
\documentstyle{amsppt}
\magnification=1200
\hoffset=-0.5pc
\nologo
\vsize=57.2truepc
\hsize=38.5truepc
\spaceskip=.5em plus.25em minus.20em
 \define\armcusgo{1}
 \define\armgojen{2}
 \define\atibottw{3}
 \define\gotaybos{4}
  \define\poisson{5}
  \define\souriau{6}
  \define\singula{7}
 \define\singulat{8}
 \define\topology{9}
  \define\smooth{10}
  \define\direct{11}

\define\singulth{13}
\define\kemneone{14}
\define\kirwaboo{15}
\define\lermonsj{16}
\define\naramntw{17}
\define\naramnth{18}
\define\seshaone{19}
\define\seshaboo{20}
\define\sjamlerm{21}
\define\weylbook{22}
\noindent{\bf PUB. IRMA, Lille - 1993\newline\noindent
Vol.33, N$^{\roman o}$ \, XI}\newline\noindent
hep-th/9312113

\topmatter
\title Poisson geometry of flat connections\\
for SU(2)-bundles on surfaces
\endtitle
\author Johannes Huebschmann
\endauthor
\affil
Universit\'e des Sciences et Technologie
de Lille
\\
U. F. R. de Math\'ematiques
\\
F-59 655 VILLENEUVE D'ASCQ, France
\\
Huebschm\@GAT.UNIV-LILLE1.FR
\endaffil
\date{November 21, 1993}
\enddate
\abstract
{In earlier work we have shown that
the moduli space $N$ of flat connections
for the (trivial) $\roman{SU(2)}$-bundle
on a closed surface of genus $\ell \geq 2$
is
a stratified symplectic space
so that, in particular, the data give rise to a
Poisson algebra $(C^{\infty}N,\{\cdot,\cdot\})$
of continuous functions on $N$;
furthermore, the strata are K\"ahler manifolds,
and the stratification
consists of
an open connected
and dense submanifold $N_Z$ of real dimension $6(\ell-1)$,
a connected stratum $N_{(T)}$
of real dimension $2\ell$,
and $2^{2\ell}$ isolated points.
In this paper
we show that,
close to each point of
$N_{(T)}$,
the space $N$
and Poisson algebra $(C^{\infty}N,\{\cdot,\cdot\})$
look like
a product of $\bold C^{\ell}$ endowed with the standard symplectic
Poisson
structure
with the reduced
space and Poisson algebra of the
system of
$(\ell-1)$ particles in the plane with total angular momentum zero,
while  close to
one of the isolated points,
the
Poisson algebra $(C^{\infty}N,\{\cdot,\cdot\})$ looks like
that of the reduced
system of $\ell$
particles in $\bold R^3$
with total angular momentum zero.
Moreover,
in the genus two case
where the space $N$ is known to be
smooth
we locally describe
the Poisson algebra $(C^{\infty}N,\{\cdot,\cdot\})$
and
the various underlying
symplectic structures
on the strata
and their mutual positions
explicitly
in terms of
the Poisson structure.}
\endabstract

\endkeywords
\endsubjclass

\endtopmatter
\document
\rightheadtext{Poisson geometry of flat SU(2)-connections}

\beginsection Introduction

Let $\Sigma$ be a closed surface
of genus $\ell\geq 2$, write
$\pi$ for its fundamental group,
let $G =\roman {SU}(2)$,
let $N$ be
the moduli space  of flat
connections
for the trivial $\roman {SU}(2)$-bundle on $\Sigma$,
and suppose the Lie algebra $g$ of $G$
endowed with a $G$-invariant inner product.
Let
$Z=\{\pm 1\}$
denote the centre of $G$ and
$T=S^1 \subseteq G$
the standard circle subgroup inside
$G$; it is a maximal torus.
The decomposition
of $N$
according to orbit types of flat connections
looks like
$$
N = N_{G} \cup N_{(T)} \cup N_{Z}
\tag0.1
$$
where the subscript refers to the conjugacy class
of stabilizer.
In an earlier paper
{}~\cite\singulat\
we proved that this decomposition is a stratification
in the strong sense;
the stratum
$N_{Z}$
is open, connected, and dense,
and referred to as {\it top stratum\/}.
While the space $N$ is known to have
a structure of a normal projective variety \cite\seshaone,
it is unclear whether and how
the stratification (0.1)
relates
to the corresponding
complex analytic one;
I am indebted to M. S. Narasimhan
for this comment.
In another earlier paper
{}~\cite\direct,
we
proved that
the  space $N$
inherits a structure of
a {\it stratified symplectic space\/}
in the sense of {\smc Sjamaar-Lerman}~\cite\sjamlerm;
in particular,
we constructed a Poisson algebra
$(C^{\infty}N,\{\cdot,\cdot\})$
of continuous functions on $N$
which,
on each stratum, restricts to a symplectic
Poisson algebra of smooth functions
in the ordinary sense.
In the present paper
we
describe
the strata locally explicitly in terms of certain related
classical constrained systems;
in particular,
for the special case of genus two,
we
describe
the Poisson algebra
and resulting Poisson geometry
of the space $N$ explicitly.
Here is
our first result.

\proclaim{Theorem 1}
Near  a point
of the top stratum $N_{Z}$,
the space $N$ looks like
$\bold C^{3(\ell-1)}$, endowed with the standard symplectic
and hence Poisson structure.
Furthermore, near  a point
of $N_{(T)}$,
the space $N$
and Poisson algebra $(C^{\infty}N,\{\cdot,\cdot\})$
look like
the product of a copy of
$\bold C^{\ell}$, endowed with the standard symplectic
Poisson structure,
with
the reduced
reduced space and Poisson algebra of the reduced
system of
$(\ell-1)$ particles in the plane
with total angular momentum zero.
Finally, near  a point
of $N_{G}$,
the space $N$
and Poisson algebra $(C^{\infty}N,\{\cdot,\cdot\})$
look like
the
reduced space and Poisson algebra
of
the reduced system of $\ell$
particles in ordinary space
with total angular momentum zero.
\endproclaim

Our approach also yields a proof of the following
result, established by {\smc  Narasimhan and Ramanan\/}
by other methods \cite\naramntw, cf. also \cite\seshaboo.

\proclaim{Theorem 2}
As a space, $N$ is smooth
if and only if
$\Sigma$ has genus two.
\endproclaim

In fact, for genus two, $N$ is just complex projective 3-space;
however,
the algebra
$C^{\infty}N$
then does {\it not\/}
coincide with the standard one of smooth functions.
It follows that, in particular,
the reduced {\it space\/} underlying
the reduced system of two
particles in ordinary space
with total angular momentum zero
is smooth, in fact, just $\bold R^6$.
This does not seem to have been known before.
However its reduced Poisson algebra is
considerably
more complicated
than the standard one on $\bold R^6$;
see Section 7 below for details.
\smallskip
The reduced  systems
in the genus two case
coming into play in Theorem 1
will be described explicitly
in Sections 6 and 7
below.
In particular, this will show how the
symplectic
structures
on the strata behave in the genus two case.
The reduced system of a
single particle in the plane
with total angular momentum zero
has been understood for a while
\cite\gotaybos\
and
is
easy to describe:
Consider the plane $\bold R^2$,
with coordinates $x_1,x_2$.
Consider the algebra of (continuous) functions
on the plane
generated by $x_1,x_2$
and the radius function $r$;
thus these generators are subject to the relation
${
x_1^2 + x_2^2 = r^2.
}$
The assignment
$$
\{x_1,x_2\} = 2r,
\quad
\{x_1,r\} = 2 x_2,
\quad
\{x_2,r\} = -2 x_1
\tag0.2
$$
endows
this algebra with a Poisson structure
$\{\cdot,\cdot\}$;
notice away from the origin,
in suitable coordinates,
this Poisson structure
amounts to the standard symplectic
Poisson structure
while it
degenerates at the origin.
The
reduced Poisson algebra
of a
single particle in the plane
with total angular momentum zero
is that of continuous functions
in the plane
that are smooth
in the (dependent) variables $x_1,x_2,r$,
with the Poisson structure {\rm (0.2)}.
\smallskip
The reduced system of two
particles in ordinary space
with total angular momentum zero
is much more complicated.
See Section 7 below for details.
Suffice it to
mention at this stage that the reduced Poisson algebra
is generated by ten algebraic functions
on $\bold R^6$
six thereof being
algebraically
independent.
\smallskip

Our method is to
study the local models obtained in our earlier papers
\cite\singula\ -- \cite\direct\
by means of
invariant theory
and to exploit a result of
{\smc Kempf-Ness}~\cite\kemneone\ and
{\smc Kirwan}~\cite\kirwaboo,
thereby playing off against each other
invariant theory  over the reals and over the complex numbers.
To get our hands on the local models
we calculate
various group cohomology spaces
as suitable
unitary representations
explicitly in Section 2 below.
Our approach is quite general and applies to
arbitrary genus and structure group.
Its very simplicity should enable one
to study more complicated moduli spaces.
\smallskip
In another guise, the moduli space $N$ is that of semi stable
holomorphic vector bundles on $\Sigma$
(with reference to a choice of holomorphic structure)
of rank 2, degree 0, and trivial determinant.
This space and related ones have been studied extensively in the literature
\cite\naramntw,
\cite\naramnth,
\cite\seshaone.
In particular,
for genus $\ell \geq 2$,
the space
$\Cal K =N_{G} \cup N_{(T)}$
is known to be the {\it Kummer\/} variety
of $\Sigma$ associated with its Jacobien $J$
and the canonical involution thereupon.
Theorem 1 above implies
in particular
that, for genus $\geq 3$,
the Kummer variety
$\Cal K$ is precisely the singular locus of $N$,
a result due to
{\smc Narasimhan-Ramanan}~\cite\naramntw.
Theorem 1 above has the following consequence:

\proclaim{Corollary}
When $\Sigma$ has genus $\ell \geq 2$,
the
Poisson algebra $(C^{\infty}N,\{\cdot,\cdot\})$
detects the Kummer variety
$\Cal K$
in $N$
together with its
$2^{2 \ell}$ double points.
More precisely,
$\Cal K$
consists of the points of $N$ where the rank of
the Poisson structure
is not maximal, the double points
being those where the rank is zero.
\endproclaim

We now make a few comments about the special case
where
$\Sigma$ has genus two.
The space $N$ then equals complex projective
3-space
and $\Cal K$ is the Kummer surface
associated with the Jacobien of $\Sigma$, cf.
{\smc Narasimhan-Ramanan}~\cite\naramntw.
In the literature,
this case has been considered somewhat special
since as a {\it space\/}
$N$ is then actually smooth.
However, from our point of view, there is
{\it no\/}
exception.
As a stratified {\it symplectic\/} space,
$N$ still has singularities,
our algebra
$C^{\infty}N$ is {\it not\/}
that of smooth functions in the ordinary sense,
and the Kummer surface $\Cal K$
is the complement of the top stratum $N_Z$
and hence still precisely the singular locus
in the sense of stratified symplectic space;
in particular,
the symplectic structure
on the top stratum does {\it not\/}
extend to the whole space.
It is interesting to observe that the stratification
(0.1) is {\it finer\/}
than the standard complex analytic one on complex
projective 3-space.
\smallskip
I am indebted to R. Cushman
and M. S. Narasimhan
for discussions,
and to
H. P. Kraft,
R. Montgomery, and A. Weinstein for
some e-mail correspondence.

\medskip\noindent{\bf 1. The stratification}\smallskip\noindent
Consider the standard presentation
$$
\Cal P  = \big\langle x_1,y_1,\dots, x_{\ell},y_\ell;
r\big\rangle ,\quad r= \prod_{j=1}^{\ell}[x_j,y_j],
\tag1.1
$$
of $\pi$.
The choice of generators induces an embedding
of
$\roman{Hom}(\pi,\roman G)$
into
$G^{2{\ell}}$, and in this way
the former will henceforth be viewed as a subspace of the latter.
Moreover the holonomies
with reference to the chosen generators
induce a diffeomorphism
in the appropriate sense
from $N$ onto the representation space
$\roman{Rep}(\pi,G) =\roman{Hom}(\pi,G) \big / G$;
see \cite\smooth\ for details.
Henceforth we do not distinguish in notation between
$N$ and
$\roman{Rep}(\pi,G)$.
\smallskip
We now reproduce briefly the stratification
of the moduli space $N$ of
flat $\roman {SU}(2)$-connections:
The choice of $T$ induces an embedding
of
$\roman{Hom}(\pi,T)$
into $\roman{Hom}(\pi,G)$, and
the
choice of
generators identifies
$\roman{Hom}(\pi,T)$ with $T^{2{\ell}}$.
Let $Y$
be the $G$-orbit of
$\roman{Hom}(\pi,T)$
in $\roman{Hom}(\pi,G)$
under the adjoint action.
Then
$\roman{Hom}(\pi,\roman {SU}(2))$
decomposes
into
$\roman{Hom}(\pi,\roman {SU}(2))\setminus Y$
and $Y$.
Each
point in
$\roman{Hom}(\pi,\roman {SU}(2))\setminus Y$
has stabilizer the centre $Z=\{\pm 1\}$ of $G$,
in fact, these are the irreducible representations
of $\pi$ in $G$,
while
each point in
$\roman{Hom}(\pi,T) \cong T^{2{\ell}}$
has stabilizer $T$.
Furthermore the inclusion
of
$\roman{Hom}(\pi,T)$ into  $Y$
induces a bijection
of
$
\roman{Hom}(\pi,T) \big/ W
$
onto
$Y\big/ G$
where $W=\bold Z/2$
refers to the Weyl group of $\roman {SU}(2)$.
Now
$\roman{Hom}(\pi,T)$ looks like
$T^{2{\ell}}$,
and the non-trivial element $w$ of $W$
acts on
$T^{2{\ell}}$
by the assignment
of
$\left(\overline\zeta_1,\dots,\overline\zeta_{2{\ell}}\right)$
to
$\left(\zeta_1,\dots,\zeta_{2{\ell}}\right)$
where as usual \lq\lq $\zeta \mapsto \overline \zeta$\rq\rq\
refers to
complex conjugation.
Hence
the fixed point set of
the action of $W$ on
$\roman{Hom}(\pi,T)$
is
$$
N_{G}  = \{\phi; \phi(x_j) = \pm 1,\
\phi(y_j) = \pm 1,\ 1 \leq j \leq {\ell}\},
$$
and the $W$-action is free
on
$\roman{Hom}(\pi,T) \setminus N_{G}$.
Thus, cf. (0.1), the resulting decomposition of $N$ looks like
${
N = N_{G} \cup N_{(T)} \cup N_{Z};
}$
here
$N_{Z} =(\roman{Hom}(\pi,G)\setminus Y)\big/G$
and
$N_{(T)} =(Y \setminus N_{G})\big/G$.
In \cite\singulat\ this
decomposition has been proved to be a stratification
in the strong sense.
Notice the
projection map
from
$\roman{Hom}(\pi,G)\setminus Y$
onto $N_{Z}$
is actually a principal $\roman {SO}(3,\bold R)$-bundle map,
whence
$N_{Z}$
is a smooth manifold
of real dimension $6({\ell}-1)$.
Furthermore
$N_{G}$ consists of $2^{2{\ell}}$ isolated points while
$N_{(T)}$ is manifestly a smooth connected manifold
of real dimension $2{\ell}$;
in fact,
the projection map
\linebreak
$\roman{Hom}(\pi,T) \to
N_{(T)}$
is a branched 2-fold covering, branched
at
the points of $N_{G}$.
\smallskip
We conclude with the remark that
$\roman{Hom}(\pi,T)$
is the group of characters of $\pi$ and hence
may be canonically identified with the Jacobien of $\Sigma$.
The quotient
$\roman{Hom}(\pi,T) \big/ W$
is then the Kummer variety mentioned in the Introduction.

\beginsection 2. The local model

Let $\phi \colon \pi \to G$ be a homomorphism,
and consider the group cohomology
$\roman H^*(\pi,g_{\phi})$.
The Lie bracket on $g$ induces a graded Lie bracket
$$
[\cdot,\cdot]_{\phi}
\colon
\roman H^*(\pi,g_{\phi})
\otimes
\roman H^*(\pi,g_{\phi})
@>>>
\roman H^*(\pi,g_{\phi})
\tag2.1
$$
on $\roman H^*(\pi,g_{\phi})$.
Further,
after a choice of fundamental class
in $\roman H_2(\pi,\bold Z)$
has been made,
the orthogonal structure on $g$ induces a
non-degenerate bilinear pairing
$$
(\cdot,\cdot)_{\phi}
\colon
\roman H^j(\pi,g_{\phi})
\otimes
\roman H^{2-j}(\pi,g_{\phi})
@>>>
\bold R
\tag2.2
$$
which, for $j=1$, amounts to a
{\it symplectic\/} structure
$$
\sigma_{\phi}
\colon
\roman H^1(\pi,g_{\phi})
\otimes
\roman H^1(\pi,g_{\phi})
@>>>
\bold R
\tag2.3
$$
on
$\roman H^1(\pi,g_{\phi})$.
Moreover,
$\roman H^0(\pi,g_{\phi})$
equals the Lie algebra $z_{\phi}$ of the stabilizer
$Z_{\phi} \in G$ of $\phi$,
the pairing
(2.2) identifies
$\roman H^2(\pi,g_{\phi})$
with the dual
$z^*_{\phi}$,
and  the assignment
$$
\Theta_{\phi}
\colon
\roman H^1(\pi,g_{\phi})
@>>>
\roman H^2(\pi,g_{\phi}),
\quad
\Theta_{\phi}(\eta)
=
\frac 12 [\eta,\eta]_{\phi},\quad
\eta \in \roman H^1(\pi,g_{\phi}),
\tag2.4
$$
is a momentum
mapping for the
action
of the stabilizer
$Z_{\phi}$
on
$\roman H^1(\pi,g_{\phi})$.
See \cite\singula\ for details.
\smallskip
Using the recipe
exploited several times
in our earlier papers,
we endow the Marsden-Weinstein reduced space
${
\roman H_{\phi} = \Theta_{\phi}^{-1}(0)\big/ Z_{\phi}
}$
with a smooth structure,
cf. e.~g. \cite\smooth.
Let
$V_{\phi}$ denote the zero locus
$\Theta_{\phi}^{-1}(0)$ of
$\Theta_{\phi}$ in $\roman H^1(\pi,g_{\phi})$,
write
$I_{\phi} \subseteq C^{\infty}(\roman H^1(\pi,g_{\phi}))$
for the ideal of smooth functions
on $\roman H^1(\pi,g_{\phi})$
that vanish on
$V_{\phi}$, and let
$$
C^{\infty}(\roman H_{\phi})
= \left(C^{\infty}(\roman H^1(\pi,g_{\phi}))\right)^{Z_{\phi}}
\big/ \left(I_{\phi}\right)^{Z_{\phi}},
\tag2.5
$$
the algebra of smooth $Z_{\phi}$-invariant
functions on
$\roman H^1(\pi,g_{\phi})$,
modulo the ideal
$\left(I_{\phi}\right)^{Z_{\phi}}$
of smooth $Z_{\phi}$-invariant
functions
that vanish on $V_{\phi}$.
With respect to the decomposition
into connected components of
orbit types,
(2.5)
endows
$\roman H_{\phi}$
with a smooth structure.
Moreover,
cf. ~\cite\armcusgo,
the symplectic Poisson
structure on
$\roman H^1(\pi,g_{\phi})$
passes to a Poisson structure
$\{\cdot,\cdot\}_{\phi}$ on
$C^{\infty}(\roman H_{\phi})$.
The
space
$\roman H_{\phi}$
together with the Poisson algebra
$\left(C^{\infty}(\roman H_{\phi}),\{\cdot,\cdot\}_{\phi}\right)$
is our local model for the moduli space $N$ near
the point represented by $\phi$, as a stratified symplectic space.
See \cite\smooth\  and \cite\direct\ for details.
\smallskip
For a homomorphism $\phi$
representing a point in the top stratum
$N_{Z}$, that is,
having the property that
$\roman H^0(\pi,g_{\phi})$ is zero,
the cohomology space
$\roman H^1(\pi,g_{\phi})$
is a finite dimensional
symplectic vector space
of real dimension $6({\ell}-1)$ and, near a point
represented by $\phi$,
as a
symplectic
manifold,
the moduli space
$N$ looks like
$\roman H^1(\pi,g_{\phi})$,
viewed as a symplectic vector space.
In particular the Poisson structure then
amounts to
the usual symplectic one.
\smallskip
To understand the other strata we must
at first calculate the cohomology groups
$\roman H^j(\pi,g_{\phi})$, for $j=0,1,2$,
as $Z_{\phi}$-representation spaces.
It seems wise to proceed in somewhat greater generality:
\smallskip
Let $K$ be an arbitrary compact Lie group,
let the genus ${\ell}$  of $\Sigma$ be arbitrary $\geq 2$,
and let
$M$ be a  finite dimensional
(real) orthogonal representation
of $K$
together with a compatible
$\pi$-module structure
in the sense that
for every $x \in K,\ y \in \pi,\ m \in M$,
$$
x(ym) = y (x m).
$$
The cohomology groups
$\roman H^j(\pi,M),\ j=0,1,2$,
are calculated by means of the standard chain
complex
arising from the presentation (1.1) in the usual way.
It looks like
$$
M
@>{d^0}>>
M^{2{\ell}}
@>{d^1}>>
M.
\tag2.6
$$
Now $\roman H^1(\pi,M)$
inherits a {\it Hermitian structure\/}
in the following way:
The inner product on $M$ extends to one on
$M^{2{\ell}}$
in the obvious way
and
the operation of taking orthogonal complements
yields a
canonical decomposition
$$
M^{2{\ell}}
\cong
d^0(M)
\oplus
\roman H^1(\pi,M)
\oplus
d^1(M^{2{\ell}})
\tag2.7
$$
whence in particular
$\roman H^1(\pi,M)$
inherits an inner product
from that on $M^{2{\ell}}$.
The symplectic structure arises in the same way
as
(2.3) above,
that is, it now looks like
$$
\sigma
\colon
\roman H^1(\pi,M)
\otimes
\roman H^1(\pi,M)
@>>>
\bold R ;
\tag2.8
$$
it involves a choice of fundamental class
in $\roman H_2(\pi,\bold Z)$.
The two structures
induce a complex structure $*$
and then combine to a Hermitian structure
on $\roman H^1(\pi,M)$.
Hence the
orthogonal
$K$-representation on $M$
induces a structure of
unitary
$K$-representation on $\roman H^1(\pi,M)$.
\smallskip
Until the end of this Section
we now suppose that $G$
is an arbitrary compact Lie group, with
 Lie algebra $g$ endowed with a
$G$-invariant inner product.
Let $\phi \colon \pi \to G$
be a homomorphism.
The Lie algebra $g_{\phi}$, endowed with the
resulting $\pi$-module structure, decomposes
as a $\pi$-module
into the direct sum
of the Lie algebra
$z_{\phi}= \roman H^0(\pi,g_{\phi})$
of the stabilizer $Z_{\phi}\subseteq G$ of $\phi$
and its orthogonal complement
$z_{\phi}^{\bot}$ in $g$ with reference to the
inner product on $g$.
The above remarks apply
with $K = Z_{\phi}$
and $M$ any one of
$g_{\phi}$,\,
$z_{\phi}$,\, and
$z_{\phi}^{\bot}$.
Hence
the cohomology spaces
$\roman H^1(\pi,z_{\phi})$,
$\roman H^1(\pi,g_{\phi})$,
$\roman H^1(\pi,z_{\phi}^{\bot})$,
inherit structures of a unitary $Z_{\phi}$-representation.

\proclaim{Theorem 2.9}
For $\ell \geq 2$, as a unitary $Z_{\phi}$-representation,
the space $\roman H^1(\pi,g_{\phi})$
decomposes into
a direct sum of
${\ell}$ copies of
$z_{\phi}\otimes \bold C$
and
${\ell}-1$ copies of
$z_{\phi}^{\bot}\otimes \bold C$,
both
$z_{\phi}\otimes \bold C$
and
$z_{\phi}^{\bot}\otimes \bold C$
being viewed as unitary $Z_{\phi}$-representations
in the obvious way.
\endproclaim

\proclaim{Corollary 2.10}
Under the circumstances of the Theorem,
as far as the resulting symplectic structure
is concerned,
viewed as an affine symplectic manifold
with a Hamiltonian $Z_{\phi}$-action, the space
$\roman H^1(\pi,g_{\phi})$
decomposes into a product
$$
(z_{\phi}^{\ell})\times (z_{\phi}^{\bot})^{{\ell}-1}
\times
\,i\,\left((z_{\phi}^{\ell})\times (z_{\phi}^{\bot})^{{\ell}-1}\right)
$$
of two Lagrangian
subspaces,
and hence looks like a cotangent bundle.
\endproclaim

To prove the theorem we obsere first that
$z_{\phi}$
is a trivial $\pi$-module in such a way that
$z_{\phi}= \roman H^0(\pi,z_{\phi})$
whence the canonical map from
$\roman H^2(\pi,z_{\phi})$
to $\roman H^2(\pi,g_{\phi})$
is an isomorphism and, furthermore,
$\roman H^0(\pi,z_{\phi}^{\bot})$
and
$\roman H^0(\pi,z_{\phi}^{\bot})$ are both zero.
Consequently the standard chain complex computing
$\roman H^*(\pi,g_{\phi})$
decomposes into
the two chain complexes
$$
z_{\phi}
@>{d^0}>>
z_{\phi}^{2{\ell}}
@>{d^1}>>
z_{\phi},\qquad
z_{\phi}^{\bot}
@>{d^0}>>
(z_{\phi}^{\bot})^{2{\ell}}
@>{d^1}>>
z_{\phi}^{\bot},
$$
the former having both $d^0$ and $d^1$
zero while the latter
having
$d^0$ injective and $d^1$ surjective.
In particular,
as a symplectic,
in fact Hermitian vector space,
$\roman H^1(\pi,g_{\phi})$
decomposes into the direct sum of
$\roman H^1(\pi,z_{\phi})$
and
$\roman H^1(\pi,z_{\phi}^{\bot})$
in such a way that (i)
$\roman H^1(\pi,z_{\phi})$
amounts to a direct sum of
$2{\ell}$ copies of
$z_{\phi}$
and (ii) the real dimension of
$\roman H^1(\pi,z_{\phi}^{\bot})$
equals $2(\ell-1)$ times the real dimension
of $z_{\phi}^{\bot}$.
To complete the argument for the theorem it will therefore suffice to prove
the following.

\proclaim{Lemma 2.11}
Suppose that
$\roman H^0(\pi,M)$ is zero.
Then, as a real unitary
$K$-representa-
\linebreak
tion,
$\roman H^1(\pi,M)$
is isomorphic to a direct sum of
${\ell}-1$ copies of
$M\otimes C$
with the obvious structure of unitary $K$-representation.
\endproclaim

\demo{Proof}
Since
$\roman H^0(\pi,M)$ is zero
the canonical decomposition (2.7) of real orthogonal
$K$-representations
implies at once that,
as
real orthogonal
$K$-representations,
$\roman H^1(\pi,M)$ and
$M^{2({\ell}-1)}$ are isomorphic.
However
the complex,
euclidean and symplectic structures
respectively $*$, $\cdot$\ , and $\sigma$, on
$\roman H^1(\pi,M)$
are related by
$$
u\cdot v = \sigma (u,*v),\quad
u,v \in \roman H^1(\pi,M).
$$
Hence as a hermitian space,
$\roman H^1(\pi,M)$ is isomorphic to
a direct sum of $\ell-1$ copies
of $M \oplus iM = M \otimes \bold C$. \qed
\enddemo

\beginsection 3. The symplectic links of the points in the middle stratum

The purpose of this Section is to prove the following.

\proclaim{Lemma 3.1}
Topologically,
the points of
$N_{(T)}$
are non-singular in $N$ if and only if
$\Sigma$ has genus $\ell=2$.
Consequently the moduli space $N$ is smooth only if
$\Sigma$ has genus $\ell=2$.
\endproclaim

As a $T$-module,
$g = t \oplus t^{\bot}$,
the direct sum of the Lie algebra $t\cong \bold R$ of $T$
and its orthogonal complement
$t^{\bot} \cong \bold R^2$,
and, by construction, $T$ acts on
$t^{\bot} \cong \bold R^2$
through the 2-fold covering map
$T \to \roman {SO}(2,\bold R)$,
the group $\roman {SO}(2,\bold R)$
being identified with that of rotations
of
$t^{\bot} \cong \bold R^2$.
In view of what was said in the previous Section,
as a unitary $S^1$-representation,
the space
$\roman H^1(\pi,g_{\phi})$
decomposes into a direct sum
$$
\roman H^1(\pi,g_{\phi})
\cong
(t\otimes \bold C)^{{\ell}}  \oplus (t^{\bot}\otimes \bold C)^{{\ell}-1},
\tag3.2
$$
that on
$(t\otimes \bold C)^{{\ell}}$
is manifestly trivial, and we are left
with the unitary
$S^1$-representation
on
$(t^{\bot}\otimes \bold C)^{{\ell}-1}$
and its momentum mapping.
By standard principles the momentum mapping is determined
by the representation.
Clearly it will suffice to examine
the corresponding
$\roman {SO}(2,\bold R)$-representation.
\smallskip
On a single copy of
$$
\bold C^2 =
t^{\bot} \oplus \, i \, t^{\bot} = \bold R^2 \oplus \, i \, \bold R^2,
\tag3.3
$$
the requisite unitary
$\roman {SO}(2,\bold R)$-representation
amounts to rotation on the real and imaginary summands
in the decomposition (3.3).
However,
under the canonical identification of
$\roman {SO}(2,\bold R)$ with
the group $\roman U(1)$,
in a suitable basis of $\bold C^2$,
the representation
is given by the assignment of
$\left[
\matrix
\zeta &0\\
0 & \overline \zeta
\endmatrix
\right]
$
to $\zeta \in \roman U(1)$.
Hence,
with the notation $n=\ell-1$,
the resulting unitary
representation
of $\roman {SO}(2,\bold R)$ on
${
\roman H^1(\pi,t^{\bot})
\cong
\left(\bold C^2\right)^{n}
}$
looks like
$$
\zeta \longmapsto
\left(
\left[
\matrix
\zeta &0\\
0 & \overline \zeta
\endmatrix
\right],
\dots,
\left[
\matrix
\zeta &0\\
0 & \overline \zeta
\endmatrix
\right]
\right) ,
\qquad
\text{for}\quad \zeta \in \roman U(1).
$$
Its momentum mapping
${
\mu \colon
\left(\bold C^2\right)^{n}
@>>>
\bold R
}$
is given by
$$
\mu(w_1,z_1,\dots,w_{n},z_{n})
=
\frac 12
\left (|z_1|^2+\dots +|z_{n}|^2 -(|w_1|^2+\dots+ |w_{n}|^2)\right)
$$
whence the zero locus
$
\mu^{-1}(0)
$
consists of the $2n$-tuples
${
(w_1,z_1,\dots,w_{n},z_{n})
\in
\left(\bold C^2\right)^{n}
}$
satisfying the equation
$$
|z_1|^2+\dots +|z_{n}|^2 = |w_1|^2+\dots+ |w_{n}|^2.
$$
Consequently the intersection
of the zero locus
$
\mu^{-1}(0)
$
with the unit sphere
\linebreak
$S^{4n-1} \subseteq \left(\bold C^2\right)^{n}$
is the product
${
S^{2n-1} \times S^{2n-1}
}$
of two spheres of radius $\frac 12$,
and the circle group $S^1$ acts diagonally thereupon.
Thus the symplectic link $L$
looks like
${
L= S^{2n-1} \times_{S^1} S^{2n-1}
}$
and, moreover, fits into
a fibre bundle
over
$P_{n-1}\bold C \times P_{n-1}\bold C$,
with fibre a circle.
In particular we see that $L$ is a sphere, in fact a circle,
if and only if $n=1$, that is, $\ell=2$.
Thus, topologically,
the points of
$N_{(T)}$
are non-singular in $N$ if and only if
the genus of $\Sigma$ equals two.
This proves the Lemma.

\beginsection 4. Symplectic and categorical quotients

We reproduce a special case
of a result due to
{\smc Kempf-Ness} and
{\smc Kirwan}.
\smallskip
Let $E$ be a finite dimensional unitary representation
of a compact Lie group $K$.
The $K$-action on $E$ extends to
a complex representation of
its complexification
$K^{\bold C}$.
Associated with it is the well known unique momentum mapping
${
\mu \colon E @>>> k^*
}$
vanishing at the origin
which, after a choice of basis
so that $E \cong \bold C^m$,
looks like
$$
\mu^{x}(\bold z) = \frac i2 \sum_{j,k}x_{j,k}\overline z_j z_k,
\quad
\text{where}
\quad
\bold z = (z_1,\dots,z_m) \in E,\quad x=\left[x_{j,k}\right] \in k.
\tag4.1
$$

\proclaim{Lemma 4.2}
The canonical map
$
E\big/\big/K @>>>
E\big /\big/K^{\bold C}
$
from
the symplectic quotient
${
E\big/\big/K = \mu^{-1}(0)\big/K
}$
to the (affine) categorical quotient
$E\big /\big/K^{\bold C}$
induced by the inclusion of
$\mu^{-1}(0)$ into $E$
is a homeomorphism.
\endproclaim

\demo{Proof}
Injectivity
is established in
\cite\kirwaboo\ (7.2)
while the surjectivity may be found in
\cite\kemneone. \qed
\enddemo

\medskip\noindent{\bf 5. Some classical invariant theory}
\smallskip\noindent
We shall determine
the requisite categorical quotients
by means of classical invariant theory.
We briefly reproduce the necessary results
from {\smc Weyl}~\cite\weylbook,
simultaneously
over $\bold R$ and
$\bold C$,
and we write
$\bold K$ for either ground field,
to have a neutral notation:

\proclaim{5.1}
Let $n \geq 2$ and ${\ell} \geq 1$.
The following is a complete list of {\it invariants\/} for the
$\roman{SO}(n,\bold K)$-action on $(\bold K^n)^{\ell}$:
\newline\noindent
{\rm (5.1.1)} the $\frac {{\ell}({\ell}+1)}2$ inner products $u_j u_k$, where
   $u_j,u_k \in \bold K^n, 1 \leq j \leq k\leq \ell$;
\newline\noindent
{\rm (5.1.2)} the $\binom {\ell}n$ determinants
$|u_{j_1} u_{j_2} \dots u_{j_n}|$
where $u_{j_1}, u_{j_2}, \dots ,u_{j_n} \in \bold K^n$,
\linebreak $1 \leq j_1 <j_2 <\dots< j_n\leq \ell$;
these, of course, may be non-zero only if ${\ell} \geq n$.
\newline\noindent
When ${\ell}<n$ these
invariants are independent.
When ${\ell}\geq n$, there are
\newline\noindent
--- $\frac{\binom {\ell}n\left(\binom {\ell}n +1\right)}2$
relations of the kind
$$
|u_{j_1} u_{j_2} \dots u_{j_n}|
|v_{j_1} v_{j_2} \dots v_{j_n}|
=
\left|
\matrix
u_{j_1} v_{j_1} & u_{j_1} v_{j_2} & \cdots& u_{j_1} v_{j_n}
\\
u_{j_2} v_{j_1} & u_{j_2} v_{j_2} & \cdots& u_{j_2} v_{j_n}
\\
\cdot           & \cdot           & \cdots& \cdot
\\
\cdot           & \cdot           & \cdots& \cdot
\\
\cdot           & \cdot           & \cdots& \cdot
\\
u_{j_n} v_{j_1} & u_{j_n} v_{j_2} & \cdots& u_{j_n} v_{j_n}
\endmatrix
\right|
\tag5.1.3
$$
and,
when ${\ell}> n$, there are
\newline\noindent
--- $\binom {\ell}{n+1}^2$ additional relations of the kind
$$
\left|
\matrix
u_{j_0} v_{j_0} &u_{j_0} v_{j_1}  & \cdots& u_{j_0} v_{j_n}
\\
u_{j_1} v_{j_0} &u_{j_1} v_{j_1}  & \cdots& u_{j_1} v_{j_n}
\\
u_{j_2} v_{j_0} &u_{j_2} v_{j_1}  & \cdots& u_{j_2} v_{j_n}
\\
\cdot           & \cdot           & \cdots& \cdot
\\
\cdot           & \cdot           & \cdots& \cdot
\\
\cdot           & \cdot           & \cdots& \cdot
\\
u_{j_n} v_{j_0} & u_{j_n} v_{j_1} & \cdots& u_{j_n} v_{j_n}
\endmatrix
\right|
=0.
\tag5.1.4
$$
These relations constitute a complete set.
\endproclaim

To spell out a crucial consequence thereof for our purposes,
write $\roman S^2(\bold K^{\ell})$
for the second symmetric power
of $\bold K^{\ell}$
and
$\Lambda^n (\bold K^{\ell})$
for the $n$'th exterior power of
$\bold K^{\ell}$.
Moreover, we denote by
$$
J \colon
(\bold C^n)^{\ell}
@>>>
\roman{so}(n,\bold R)^*
\tag5.2
$$
the momentum mapping
for the diagonal action
of $\roman{SO}(n,\bold R)$ on
$(\bold C^n)^{\ell}$,
with its canonical symplectic structure.

\proclaim{Theorem 5.3}
For $n \geq 2$ and ${\ell} \geq 1$, the  invariants
{\rm (5.1.1)} and
{\rm (5.1.2)}
induce an
$\roman{SO}(n,\bold C)$-invariant
map
$$
(\bold C^n)^{\ell}
@>>>
\roman S^2(\bold C^{\ell})
\times \Lambda^n (\bold C^{\ell})
\tag5.3.1
$$
which induces
an embedding
$$
J^{-1}(0)\big/\roman{SO}(n,\bold R)
@>>>
\roman S^2(\bold C^{\ell})
\times \Lambda^n (\bold C^{\ell})
\tag5.3.2
$$
of the
reduced space
$J^{-1}(0)\big/\roman{SO}(n,\bold R)$
into
$\roman S^2(\bold C^{\ell})
\times \Lambda^n (\bold C^{\ell})$
as the complex affine algebraic set
described by the
equations {\rm (5.1.3)} and {\rm (5.1.4)}.
\endproclaim

\demo{Proof}
In fact,
(5.3.1)
induces an embedding
of the
categorical quotient
\linebreak
$
(\bold C^n)^{\ell}\big /\big/\roman{SO}(n,\bold C)
$
into
$\roman S^2(\bold C^{\ell})
\times \Lambda^n (\bold C^{\ell})$
as the complex affine algebraic set
described by the
equations {\rm (5.1.3)} and {\rm (5.1.4)}.
By (4.3),
the canonical map identifies the symplectic quotient
with the categorical one. \qed
\enddemo

\medskip\noindent{\bf 6. The middle stratum}
\smallskip\noindent
Consider  a homomorphism $\phi\colon \pi \to G$ representing a point
of
$N_{(T)}$.
{}From (3.2) we know already that,
as a $T$-representation,
$
\roman H^1(\pi,g_{\phi})
$
decomposes
into a direct sum of
$(t\otimes \bold C)^{{\ell}}$
and
$(t^{\bot}\otimes \bold C)^{{\ell}-1}$
in such a way  that the $T$-representation on
$(t\otimes \bold C)^{{\ell}}$
is trivial.
Moreover
$t^{\bot}$ is just the ordinary plane $\bold R^2$,
and the circle group $T$ acts on it through
a 2-fold covering map onto
the group
$\roman{SO}(2,\bold R)$.
The $\roman{SO}(2,\bold R)$-representation
on $(t^{\bot}\otimes \bold C)^{{\ell}-1}$,
in turn,
amounts to
the {\it classical constrained system\/} of
${\ell-1}$ particles moving in the plane with total angular
momentum zero.
This proves the corresponding assertion in Theorem 1.
With the notation $\bold C^2 = \bold R^2 \oplus \, i \, \bold R^2$,
the
momentum mapping
of this system
looks like
$$
\mu \colon (\bold C^2)^{\ell-1} @>>> \bold R,
\quad
\mu (q_1 + ip_1,\dots,q_{\ell-1} + ip_{\ell-1})
= |q_1\,p_1| +\dots +|q_{\ell-1}\, p_{\ell-1}|.
$$
Furthermore, the extension of
the $\roman{SO}(2,\bold R)$-representation
to the complexification
$\roman{SO}(2,\bold C)$
of
$\roman{SO}(2,\bold R)$
is just the standard diagonal action of
$\roman{SO}(2,\bold C)$ on ${\ell}-1$ copies
of $\bold C^2$,
and hence the
embedding
(5.3.2)
identifies the reduced space
as a complex affine algebraic subset
of
$\roman S^2(\bold C^{\ell -1}) \times \Lambda^2 (\bold C^{\ell -1})$.
\smallskip
We now examine the special case of genus $\ell = 2$.
In this case, there is a single
$\roman{SO}(2,\bold C)$-invariant of the kind (5.1.1);
with the notation
$$
w = q + ip \in \bold R^2 \oplus \, i \, \bold R^2 =\bold C^2,
$$
it looks like
$$
ww = (q + ip)(q + ip) = qq -pp +2iqp.
$$
The holomorphic map
from
$\bold C^2$
to $\bold C$
induced by this invariant,
cf. (5.3.1),
assigns $ww\in \bold C$ to $w \in \bold C^2$
and identifies
the reduced space
$\mu^{-1}(0)\big / \roman{SO}(2,\bold R)$
with a copy of one-dimensional complex affine space $\bold C$,
that is, with the usual real affine plane $\bold R^2$.
In particular, we see
once more that, when the genus equals two,
near a point of $N_{(T)}$,
the moduli space $N$ is smooth;
however we cannot here conclude the converse also,
that is, that $N$ is smooth
near a point of $N_{(T)}$ only if the genus equals two.
\smallskip
We now determine
the reduced Poisson algebra
near a point of $N_{(T)}$
and make a few comments about the resulting Poisson geometry.
Working with polynomial functions instead of smooth functions,
following the  construction (2.5),
we must take the
$\roman{SO}(2,\bold R)$-invariants
on $\bold R^4 =\bold R^2 \times \bold R^2$
and thereafter divide out the appropriate ideal.
We continue to write
$w=(q,p) \in \bold R^2 \times \bold R^2$.
By (5.1), the four $\roman{SO}(2,\bold R)$-invariants
$$
qq, \ pp,\  qp,\  |q\,p|,
$$
constitute a complete set;
however since
the invariant
$|q\,p|$
vanishes on the zero locus,
the three invariants
$qq,\ pp,\  qp$
already generate
the reduced Poisson algebra.
Let
$$
x_1 =qq -pp,\quad
x_2 = 2qp,
\quad
r = qq + pp.
$$
It is clear that
these three functions
also constitute a set of generators.
Its advantage is that $x_1$ and $x_2$
are the
obvious coordinate functions
on the reduced space since, with the above notation,
$$
ww = x_1 + i x_2.
$$
Moreover,  the zero locus $\mu^{-1}(0)$ is defined by the equation
$
|q\,p|=0
$
whence on the reduced space we have
$$
0 = |q\,p|^2 = (qq)(pp) -(qp)^2
$$
that is,
$$
x_2^2 = 4(qp)^2= 4 (qq)(pp) = (qq + pp)^2 -(qq -pp)^2 = r^2 - x_1^2;
$$
in other words, the third generator $r$
amounts to
the usual {\it radius\/} function
in the plane with coordinates $x_1,x_2$.
Thus the reduced Poisson algebra
has generators
$x_1,x_2,r$, subject to the relation
${
x_1^2 + x_2^2 = r^2;
}$
moreover, still by construction, the Poisson brackets
among the generators are calculated
in the
{\it unreduced\/} Poisson algebra,
that is, in the algebra
of polynomials in the indeterminates
$q_1,q_2,p_1,p_2$, with the canonical symplectic Poisson
brackets $\{q_j,p_k\} = \delta_{j,k}$ etc.
This yields
$$
\{x_1,x_2\} = 2r,
\quad
\{x_1,r\} = 2 x_2,
\quad
\{x_2,r\} = -2 x_1.
$$
Thus the reduced Poisson algebra
is symplectic
everywhere except at the origin
and we see that,
symplectically,
or, more appropriately, in the {\it Poisson\/} sense,
near  a point $Q$
of $N_{(T)}$,
the space $N$ looks like
the product of a copy of
$\bold R^4$, endowed with the standard symplectic
or Poisson structure,
with a copy of
$\bold R^2$, endowed
with the above Poisson structure.
The copy
$\bold R^4$
corresponds to the stratum
$N_{(T)}$.
This shows
in particular
how
the
symplectic structure on the top stratum
degenerates
at the points of $N_{(T)}$.
Notice however that,
at a point of $N_{(T)}$,
the function $r$
is genuinely no longer smooth
in the usual sense.
To describe the {\it mutual positions\/}
of the symplectic
structures on the
strata,
we are
thus
forced to
go {\it beyond\/}
the usual {\it smooth setting\/}
and have to admit continuous functions which
are no longer smooth everywhere.
\smallskip
The {\it complexified\/}
reduced Poisson algebra
admits the following
somewhat simpler description:
It is generated by the three functions
$ww, \overline w \overline w,
w \overline w $,
with Poisson brackets
$$
\{ww,\overline w \overline w\}
= -8i w\overline w,
\qquad
\{ww, w \overline w\}
= -4i w w,
\qquad
\{\overline w \overline w, w \overline w\}
= -4i \overline w \overline w .
$$
Notice $w \overline w$ coincides with the above radius function $r$.
\smallskip
For genus $\ell >2$,
we can still take the coordinate
functions prescribed by the
corresponding embedding
given by
the {\it real\/} $\roman{SO}(2,\bold R)$-invariants
and calculate their Poisson brackets.
This yields the reduced Poisson algebra
in the general case.
We refrain from spelling out the details.

\medskip\noindent{\bf 7. The isolated points}\smallskip\noindent
To begin with,
consider a surface $\Sigma$
of arbitrary genus $\ell\geq 2$.
Let $\phi \colon \pi \to G$
be a homomorphism whose values lie in the centre $Z \subseteq G$,
so that $\phi$ represents one of the isolated points
of $N_{G}$.
By (2.9),
$\roman H^1(\pi,g)$ and
$(g\otimes \bold C)^{{\ell}}$
are isomorphic
as real unitary $G$-representations.
Hence the $G$-action
on $\roman H^1(\pi,g)$
amounts to the standard diagonal action
on ${\ell}$ copies
of $\bold R^3 \oplus \,i\,\bold R^3= \bold C^3$,
with $G$ acting on each copy of
$\bold R^3$ through
the 2-fold covering
$G \to \roman{SO}(3,\bold R)$.
This is just
the {\it classical constrained system\/} of
${\ell}$ particles moving in space with total angular
momentum zero,
except that its group is
$\roman {SU}(2)$
and acts through
the 2-fold covering
$\roman {SU}(2) \to \roman{SO}(3,\bold R)$.
This establishes the corresponding assertion in Theorem 1.
In particular,
with the notation
$$
q+ip \in \bold C^3 = \bold R^3 \oplus \,i\,\bold R^3,
$$
the
momentum mapping $\Theta$
and reduced space $\roman H_{\phi}$
look like
$$
\align
\Theta
(q_1 + i p_1,\dots,q_{\ell} +i p_{\ell})
&=
q_1 \wedge p_1
+ \dots +
q_{\ell} \wedge p_{\ell},
\tag7.1
\\
\roman H_{\phi}=
\Theta^{-1}(0)\big /\roman{SU}(2)
&=
\Theta^{-1}(0)\big /\roman{SO}(3,\bold R).
\tag7.2
\endalign
$$
Furthermore, the extension of
the $\roman{SO}(3,\bold R)$-representation
to the complexification
$\roman{SO}(3,\bold C)$
of
$\roman{SO}(3,\bold R)$
is just the standard diagonal action of
$\roman{SO}(3,\bold C)$ on ${\ell}$ copies
of $\bold C^3$,
and
(5.3.2)
identifies $\roman H_{\phi}$
as a complex affine algebraic subset
of
$\roman S^2(\bold C^{\ell}) \times \Lambda^3 (\bold C^{\ell})$.
\smallskip
How does the middle stratum $N_{(T)}$ look like in this model
of $\roman H_{\phi}$?

\proclaim {Proposition 7.3}
The middle stratum
arises from the subspace
of the zero locus $\Theta^{-1}(0)$
consisting of $\ell$-tuples
$$
(\lambda_1 w, \dots,\lambda_{\ell} w)
\in \Theta^{-1}(0) \subseteq (\bold C^3)^{\ell},
\quad
w \in \bold C^3,
\quad
\lambda_1,\dots, \lambda_{\ell} \in \bold C,
$$
and hence corresponds to the image thereof
in
$\roman S^2(\bold C^{\ell}) \times \Lambda^3 (\bold C^{\ell})$
under {\rm (5.3.2)}.
Thus in the constrained system picture,
the middle stratum
corresponds to states different from the origin
where each one of the $\ell$ particles
individually has angular momentum zero.
\endproclaim

\demo{Proof}
By construction,
the elements of
$N_{(T)}$
are represented by points
in $(\bold C^3)^{\ell}$
having stabilizer conjugate to $T$ and hence
by $\ell$-tuples of the kind
$$
\align
(w_1,\dots, w_\ell) &=
(\psi_1 v+i\rho_1 v, \dots, \psi_\ell v+i\rho_\ell v)
\\
&=
((\psi_1 +i\rho_1) v, \dots, (\psi_\ell +i\rho_\ell) v)
\in \Theta^{-1}(0) \subseteq (\bold C^3)^{\ell},
\endalign
$$
where
$v \in \bold R^3,\,\psi_j,\,\rho_j \in \bold R$.
It is clear that these
$\ell$-tuples $(w_1,\dots, w_\ell)$
are of the asserted kind.
Conversely,
let
$$
(w_1, \dots,w_{\ell})=
(\lambda_1 w, \dots,\lambda_{\ell} w)
\in \Theta^{-1}(0) \setminus \{0\}
$$
so that, for $1 \leq j \leq \ell$,
with the notation
$\lambda_j =u_j+iv_j$,
$$
w_j = \lambda_j w
=(u_j+iv_j)(q + ip) = u_jq -v_j p + i(v_jq + u_jp)
= q_j + i p_j
\tag7.4
$$
and
${
q_j \wedge p_j = \lambda_j \overline \lambda_j q \wedge p.
}$
However  the constraint
equation
$$
\Theta(q_1+ip_1,\dots,q_\ell+ip_\ell) = 0
$$
yields
$$
0=
q_1 \wedge p_1
+ \dots +
q_\ell \wedge p_\ell
=
(\lambda_1 \overline \lambda_1 + \dots+ \lambda_\ell\overline \lambda_\ell)
q\wedge p,
$$
that is
$p$ is proportional to $q$
and hence
each $p_j$ to $q_j$;
in view of
(7.4),
each $q_j$ is also proportional to $q$
and
each $p_j$ to $p$,
that is, the $2\ell$-tuple
$
(q_1,p_1,\dots q_\ell,p_\ell)
$
looks like
$$
(q_1,p_1,\dots q_\ell,p_\ell)
= (\psi_1 v,\rho_1 v, \dots, \psi_\ell v,\rho_\ell v),
\quad
v \in \bold R^3,
\quad \psi_j,\, \rho_j \in \bold R,
$$
and hence has stabilizer conjugate to $T$. \qed
\enddemo

\smallskip
We now examine the special case
of genus ${\ell}=2$ in detail.
By (5.1),
the three
holomorphic invariants
$$
w_1 w_1,\ w_1 w_2,\ w_2 w_2
\tag7.5
$$
constitute
a complete set of
invariants
for the
$\roman{SO}(3,\bold C)$-action on $(\bold C^3)^2$,
and these are independent.
Hence the categorical quotient is just a three-dimension complex affine
space and,
by virtue of (5.3), the
$\roman{SO}(3,\bold C)$-invariant holomorphic map
$$
(\bold C^3)^2
@>>>
\bold C^3,
\quad
(w_1,w_2)
\longmapsto
(w_1 w_1, w_1 w_2, w_2 w_2),
\tag7.6
$$
cf. (5.3.1),
induces a homeomorphism
$$
\roman H_{\phi}=\Theta^{-1}(0)\big /\roman{SO}(3,\bold R)
@>>>
\bold C^3.
\tag7.7
$$
In particular, this shows that, in genus two,
{\it near any of the\/} 16 {\it isolated points
of\/} $N_G$,
{\it the moduli space $N$ is smooth\/},
that is, the proof of Theorem 2 is now complete.

\proclaim{Proposition 7.8}
For genus two,
in the model $\bold C^3$ for the reduced space
$\roman H_{\phi}$,
the middle stratum  $N_{(T)}$
corresponds to the
complex quadric
$$
Q = \left\{(x,y,z) \in \bold C^3;\  y^2 = xz\right\}
\tag7.8.1
$$
with the origin removed.
\endproclaim

\demo{Proof} This is immediate from (7.4).
In fact, under (7.6),
the subspace
consisting of pairs
$
(\lambda_1 w,\lambda_2 w)
$
is mapped onto the asserted quadric. \qed
\enddemo

\smallskip
We now determine
the corresponding reduced Poisson algebra.
Working with polynomial functions instead of smooth functions,
following the  construction (2.5),
we must take the
$\roman{SO}(3,\bold R)$-invariants
on $(\bold R^3)^4$
and thereafter divide out the appropriate ideal.
We continue to write
$
(q_1,p_1,q_2,p_2) \in (\bold R^3)^4.
$
In view of (5.1) above,
the ten distinct invariants
$$
q_iq_j,\, q_ip_j,\,p_ip_j,\ 1 \leq i ,j \leq 2,
\tag7.9
$$
among
the scalar products,
together with the four
determinants
$$
|q_1 p_1 q_2|,
\
|q_1 p_1 p_2|,
\
|q_1 q_2 p_2|,
\
|p_1 q_2 p_2|,
\tag7.10
$$
constitute a complete set of invariants
for the
$\roman{SO}(3,\bold R)$-action
on
$(\bold R^3)^4$.
However, for
$(q_1+ip_1,q_2+ip_2) \in \Theta^{-1}(0)$, that is,
when
$$
\Theta
(q_1+ip_1,q_2+ip_2)
=
q_1 \wedge p_1
+
q_2 \wedge p_2 = 0,
$$
any three of
$(q_1,p_1,q_2,p_2)$
are linearly dependent, whence
the
four determinants (7.10) vanish
on the zero locus $\Theta^{-1}(0)$, and the
ten scalar products (7.9)
induce an embedding
$$
\roman H_{\phi}
@>>>
\roman S^2(\bold R^4)
\tag7.11
$$
of the reduced
space
$\roman H_{\phi}$
into the $10$-dimensional real vector space
$\roman S^2(\bold R^4)$
as a real semi-algebraic set.
More details  are given in Section 12 of our paper \cite\smooth.
Hence our reduced
Poisson algebra
of continuous functions
will be generated by
the ten scalar product invariants (7.9).
It is straightforward to calculate the Poisson brackets
between these functions;
in fact,
cf. (2.5),
this is done in the unreduced symplectic Poisson algebra
generated by the coordinate functions
on
$(\bold R^3)^4$, and we
have
$$
\{q_1 q_1, q_1 p_1\} = 2q_1 q_1,
\quad
\{q_1 p_2, q_1 p_1\} = q_1 p_2,
\tag7.12
$$
etc.
\smallskip
As a {\it real semi-algebraic subset\/} of
$\roman S^2(\bold R^4)$,
the reduced space looks more complicated
than as an
{\it affine complex algebraic\/} set.
On the other hand, the
complex algebraic structure
ignores the reduced Poisson algebra
since this algebra involves {\it additional\/} functions
which
in the complex algebraic
picture are {\it not\/}
visible.
In particular,
not all of the ten generators of the reduced Poisson algebra
will be smooth
functions on $\bold C^3$ in the ordinary sense.
\smallskip\noindent
{\smc Remark 7.14.}
In (5.4) of~\cite\lermonsj\
it is asserted that, for arbitrary $\ell$, the
$\roman{SO}(3,\bold R)$-reduced space is a double branched cover
of
the
$\roman{O}(3,\bold R)$-reduced space.
Under the present circumstances,
$\ell =2$ and
the two spaces actually coincide, that is,
the space in fact coincides with its branching locus.
The reason is that
the determinant invariants (7.10)
which distinguish between the
$\roman{O}(3,\bold R)$- and
$\roman{SO}(3,\bold R)$-reduced spaces
vanish on the zero locus
of the momentum mapping which is the same for the two.
Moreover the results
of~\cite\lermonsj\
imply that, as a stratified symplectic space,
$\roman H_{\phi}$
may be identified with the closure of a certain a nilpotent orbit in
$\roman{sp}(2,\bold R)^*$.
This orbit is made precise in Section 12 of \cite\smooth.
Moreover, the ten invariants (7.9) may be viewed as a basis of
$\roman{sp}(2,\bold R)$, and the Poisson brackets (7.12)
are given by the Lie bracket of $\roman{sp}(2,\bold R)$.
\smallskip

We summarize
the resulting
{\it Poisson geometry\/}
of the moduli space $N$ near
any of the isolated points
in the following.

\proclaim{Theorem 7.15}
In the model $\bold C^3$ for the reduced space $\roman H_{\phi}$,
away from the quadric {\rm (7.8.1)},
the Poisson structure has maximal rank
equal to six,
while at a point on the quadric {\rm (7.8.1)},
the Poisson structure has rank four
except at the origin where it has rank zero.
Moreover the ten Hamiltonian
vector fields of the generators
of the Poisson algebra,
restricted
to the quadric {\rm (7.8.1)},
are tangential to the quadric
except at the origin
where they all vanish.
\endproclaim

We note that a Hamiltonian vector field is not necessarily smooth but
under the present circumstances can still
be written
in terms of suitable frames with continuous coefficients which
are smooth on each stratum.
\smallskip
The Theorem is, of course,
an immediate consequence
of properties of our local model $\roman H_{\phi}$.
However it is instructive
to verify its statement directly as
a {\it formal consequence\/}
of properties
of the
reduced Poisson algebra
of functions
on the model $\bold C^3$ of $\roman H_{\phi}$.
We now explain this.
\smallskip
With the ten
scalar product invariant generators (7.9)
for the reduced Poisson algebra
it seems hard to visualize the resulting
Poisson geometry directly.
However,
after {\it complexification\/},
the reduced Poisson algebra admits
the following much more perspicuous description:
We may take the six
$\roman{SO}(3,\bold R)$-invariants
$$
w_1 w_1,\ w_1 w_2,\ w_2 w_2,
\ \overline w_1 \overline w_1,\ \overline w_1 \overline w_2,
\ \overline w_2 \overline w_2
\tag7.16
$$
as independent
(complex) coordinate functions, and
these six invariants, together with the four
$\roman{SO}(3,\bold R)$-invariants
$$
w_1 \overline w_1,\ w_1 \overline w_2,
\ \overline w_1 w_2,
\ w_2 \overline w_2
\tag7.17
$$
generate
the
same
algebra of  continuous complex valued
functions on the reduced space
as the ten scalar product invariants (7.9),
viewed as complex valued functions.
We note that the functions (7.17) are not smooth on the
reduced space in the ordinary sense.
\smallskip
A straightforward calculation
in the unreduced Poisson algebra
yields the following Poisson brackets:
$$
\aligned
\{w_1w_1,\overline w_1 \overline w_1\}
&= -8i w_1\overline w_1
\\
\{w_1w_1,\overline w_1 \overline w_2\}
&= -4i w_1\overline w_2
\\
\{w_1w_2,\overline w_1 \overline w_1\}
&= -4i \overline w_1 w_2
\\
\{w_1w_2,\overline w_1 \overline w_2\}
&= -2i (w_1 \overline w_1 + w_2 \overline w_2)
\\
\{w_1w_2,\overline w_2 \overline w_2\}
&= -4i w_1\overline w_2
\\
\{w_2w_2,\overline w_1 \overline w_2\}
&= -4i \overline w_1 w_2
\\
\{w_2w_2,\overline w_2 \overline w_2\}
&= -8i w_2\overline w_2
\\
\{\overline w_1\overline w_1,w_1  w_2\}
&= 4i \overline w_1 w_2
\\
\{\overline w_1\overline w_2,w_1  w_1\}
&= 4i w_1 \overline w_2
\\
\{\overline w_1\overline w_2,w_1  w_2\}
&= 2i (w_1 \overline w_1 + w_2 \overline w_2)
\\
\{\overline w_1\overline w_2,w_2  w_2\}
&= 4i  \overline w_1 w_2
\\
\{\overline w_2\overline w_2,w_1  w_2\}
&= 4i   w_1 \overline w_2 ,
\endaligned
$$
still with the convention that
Poisson brackets
between coordinate functions
not spelled out explicitly are
zero.
{}From these we derive
at first
the six hamiltonian vector fields
of the six chosen coordinate functions (7.16).
With the notation
$$
\alpha = w_1 \overline w_1,
\quad
\beta = 2 w_1 \overline w_2,
\quad
\gamma = w_2 \overline w_2,
$$
in the above chosen order of coordinate functions,
the system of Hamiltonian vector fields of the
functions (7.16) is given by
the matrix
$$
A=
2i
\left[
\matrix
0 & 0 & 0 & 4 \alpha & \beta & 0
\\
0 & 0 & 0 &  \overline \beta & \alpha + \gamma & \beta
\\
0 & 0 & 0 &  0               &\overline \beta & 4 \gamma
\\
-4 \alpha & -\overline \beta & 0 & 0 & 0 & 0
\\
- \beta & -(\alpha + \gamma) & - \overline \beta & 0 & 0 & 0
\\
0       &- \beta & -4 \gamma & 0 & 0 & 0
\endmatrix
\right] .
$$
Thus the matrix has rank six if and only if
$$
\left| \matrix
4 \alpha        & \beta           & 0
\\
\overline \beta & \alpha + \gamma & \beta
\\
0               &\overline \beta  & 4 \gamma
\endmatrix
\right|
= 4(\alpha+\gamma)(4 \alpha\gamma - \beta \overline \beta)
\ne 0.
$$
However
$$
\align
\alpha+\gamma
&=
w_1  \overline w_1
+
w_2  \overline w_2
=
q_1 q_1 +
p_1 p_1 +
q_2 q_2 +
p_2 p_2,
\\
4 \alpha\gamma - \beta \overline \beta
&=
4\left(
(w_1  \overline w_1)(w_2  \overline w_2)
-
(w_1  \overline w_2)
(\overline w_1 w_2)
\right)
=4
(w_1\wedge w_2)(\overline w_1  \wedge  \overline w_2 ) .
\endalign
$$
Hence the matrix is non-singular if and only if
$$
q_1 q_1 +
p_1 p_1 +
q_2 q_2 +
p_2 p_2 \ne 0
$$
and
if
$w_1$ and $w_2$
are linearly independent
in $\bold C^3$.
In other words, the matrix has full rank
on every point of the top stratum.
Furthermore, the matrix
is zero if and only if
$$
q_1 q_1 +
p_1 p_1 +
q_2 q_2 +
p_2 p_2 =0;
$$
this corresponds to the image of the origin in the
reduced space and hence to a point in
the stratum
$N_G$.
Finally a little thought reveals
that the matrix
has rank equal to 4 if and only of
$w_1$ and $w_2$
are linearly dependent
in $\bold C^3$.
In view of (7.4),
the points having this property correspond exactly
to the remaining stratum,
given by orbit type $(T)$ examined earlier.
However,
cf. (7.8),
this corresponds to the complex quadric $y^2 = zx$.
\smallskip
To get a complete picture,
we must examine
the Poisson brackets with
the four  remaining
$\roman{SO}(3,\bold R)$-invariants
(7.17)
and their Hamiltonian vector fields.
This does not
lead to any new difficulty, and we leave the details to the reader.
This completes our discussion of the
local
Poisson geometry
of the moduli space of flat $\roman{SU}(2)$-connections.
\smallskip
For genus $\ell >2$,
we can
proceed in much the same way
and describe the
complexified reduced Poisson algebra
and corresponding Poisson geometry.
We refrain from spelling out the details here.
\smallskip
We conclude with a comment about the
geometric
significance of the Poisson structure:
In the above model
of the genus two moduli space $N$
near any of the isolated points,
the
non-singular part of the
complex quadric
$Q \subseteq \bold C^3$
given by
the equation
$y^2 = xz$
comes of course
with its standard K\"ahler
and hence symplectic structure,
and the latter gives rise to the corresponding
symplectic Poisson structure.
However this Poisson structure
does {\it not\/}
arise
from restriction
to $Q$
of the
corresponding
standard
symplectic Poisson structure
on the ambient space
$\bold C^3$.
In fact, in the language of
theoretical physics,
the quadric $Q$ is a
{\it second class constraint\/},
and to get the correct
Poisson brackets on
$Q$ one has to introduce
{\it Dirac brackets\/}
on the ambient space.
On the other hand,
in our approach,
the Poisson algebra
on $Q$
{\it does\/}
arise
from restriction
of our
Poisson algebra
on the ambient space
$\bold C^3$.
In other words,
the ideal of functions
in this algebra
that vanish on $Q$ is
a {\it Poisson ideal\/}.
Thus we see explicitly how the
Poisson structure
encapsulates
the {\it mutual positions\/}
of the symplectic structures
on the strata.

\bigskip
\centerline{References}
\medskip
\ref \no  \armcusgo
\by J. M. Arms,  R. Cushman, and M. J. Gotay
\paper  A universal reduction procedure for Hamiltonian group actions
\paperinfo in: The geometry of Hamiltonian systems, T. Ratiu, ed.
\jour MSRI Publ.
\vol 20
\pages 33--51
\yr 1991
\publ Springer
\publaddr Berlin-Heidelberg-New York-Tokyo
\endref
\ref \no  \armgojen
\by J. M. Arms, M. J. Gotay, and G. Jennings
\paper  Geometric and algebraic reduction for singular
momentum mappings
\jour Advances in Mathematics
\vol 79
\yr 1990
\pages  43--103
\endref
\ref \no  \atibottw
\by M. Atiyah and R. Bott
\paper The Yang-Mills equations over Riemann surfaces
\jour Phil. Trans. R. Soc. London  A
\vol 308
\yr 1982
\pages  523--615
\endref
\ref \no \gotaybos
\by M. J. Gotay and L. Bos
\paper Singular angular momentum mappings
\jour J. Diff. Geom.
\vol 24
\yr 1986
\pages 181--203
\endref
\ref \no \poisson
\by J. Huebschmann
\paper Poisson cohomology and quantization.
\jour
J. f\"ur die Reine und Angewandte Mathematik
\vol  408
\yr 1990
\pages 57--113
\endref
\ref \no  \souriau
\by J. Huebschmann
\paper On the quantization of Poisson algebras
\book Symplectic Geometry and Mathematical Physics
\bookinfo Actes du colloque en l'honneur de Jean-Marie Souriau,
P. Donato, C. Duval, J. Elhadad, G.M. Tuynman, eds.;
Progress in Mathematics, Vol. 99
\publ Birkh\"auser
\publaddr Boston $\cdot$ Basel $\cdot$ Berlin
\yr 1991
\pages 204--233
\endref
\ref \no \singula
\by J. Huebschmann
\paper The singularities of Yang-Mills connections
for bundles on a surface.I. The local model
\paperinfo preprint February 1992
\endref
\ref \no \singulat
\by J. Huebschmann
\paper The singularities of Yang-Mills connections
for bundles on a surface. II. The stratification
\paperinfo preprint February 1992
\endref
\ref \no \topology
\by J. Huebschmann
\paper
Holonomies of
Yang-Mills connections
for bundles on a surface with disconnected structure group
\paperinfo preprint 1992
\endref
\ref \no \smooth
\by J. Huebschmann
\paper
Smooth structures on
moduli spaces of central Yang-Mills connections
for bundles on a surface
\paperinfo preprint 1992
\endref
\ref \no \direct
\by J. Huebschmann
\paper
Poisson
structures on certain
moduli spaces
for bundles on a surface
\paperinfo preprint 1992
\endref
\ref \no \singulth
\by J. Huebschmann
\paper The singularities of Yang-Mills connections
for bundles on a surface. III. The identification of the strata
\paperinfo in preparation
\endref
\ref \no \kemneone
\by G. Kempf and L. Ness
\paper The length of vectors in representation spaces
\jour Lecture Notes in Mathematics
\vol 732
\yr 1978
\pages 233--244
\paperinfo Algebraic geometry, Copenhagen, 1978
\publ Springer
\publaddr Berlin-Heidelberg-New York
\endref
\ref \no \kirwaboo
\by F. Kirwan
\book Cohomology of quotients in symplectic and algebraic geometry
\publ Princeton University Press
\publaddr Princeton New Jersey
\yr 1984
\endref
\ref \no \lermonsj
\by E. Lerman, R. Montgomery and R. Sjamaar
\paper Examples of singular reduction
\paperinfo preprint
\endref
\ref \no \naramntw
\by M. S. Narashiman and S. Ramanan
\paper Moduli of vector bundles on a compact Riemann surface
\jour Ann. of Math.
\vol 89
\yr 1969
\pages  19--51
\endref
\ref \no \naramnth
\by M. S. Narashiman and S. Ramanan
\paper 2$\theta$-linear systems on abelian varieties
\jour Bombay colloquium
\yr 1985
\pages  415--427
\endref
\ref \no \seshaone
\by C. S. Seshadri
\paper Spaces of unitary vector bundles on a compact Riemann surface
\jour Ann. of Math.
\vol 85
\yr 1967
\pages 303--336
\endref
\ref \no \seshaboo
\by C. S. Seshadri
\book Fibr\'es vectoriels sur les courbes alg\'ebriques
\bookinfo Ast\'erisque Vol. 96
\publ Soc. Math. de France
\yr 1982
\endref
\ref \no \sjamlerm
\by R. Sjamaar and E. Lerman
\paper Stratified symplectic spaces and reduction
\jour Ann. of Math.
\vol 134
\yr 1991
\pages 375--422
\endref
\ref \no \weylbook
\by H. Weyl
\book The classical groups
\publ Princeton University  Press
\publaddr Princeton NJ
\yr 1946
\endref
\enddocument